# Confinement of relativistic electrons in a magnetic mirror en route to a magnetized relativistic pair plasma

J. von der Linden, G. Fiksel, J. Peebles, et al.

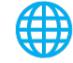 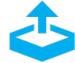

### ARTICLES YOU MAY BE INTERESTED IN

Magnetically collimated relativistic charge-neutral electron–positron beams from high-power lasers
Physics of Plasmas **28**, 074501 (2021); https://doi.org/10.1063/5.0053557

Autoresonance phenomenon in a long mirror
Physics of Plasmas **28**, 092507 (2021); https://doi.org/10.1063/5.0056531

Langmuir oscillations breaking in inhomogeneous plasma
Physics of Plasmas **28**, 092304 (2021); https://doi.org/10.1063/5.0058978

# Confinement of relativistic electrons in a magnetic mirror en route to a magnetized relativistic pair plasma

J. von der Linden,[1,2,a)] G. Fiksel,[3] J. Peebles,[4] M. R. Edwards,[1] L. Willingale,[3] A. Link,[1] D. Mastrosimone,[4] and Hui Chen[1]

### AFFILIATIONS

[1]Lawrence Livermore National Laboratory, Livermore, California 94550, USA
[2]Division E4, Max Planck Institute for Plasma Physics, 17491 Greifswald and 85748 Garching, Germany
[3]The Gérard Mourou Center for Ultrafast Optical Science, University of Michigan, Ann Arbor, Michigan 48109, USA
[4]Laboratory for Laser Energetics, University of Rochester, Rochester, New York 14623, USA

[a)]Author to whom correspondence should be addressed: jens.von.der.linden@ipp.mpg.de

### ABSTRACT

Creating a magnetized relativistic pair plasma in the laboratory would enable the exploration of unique plasma physics relevant to some of the most energetic events in the universe. As a step toward a laboratory pair plasma, we have demonstrated an effective confinement of multi-MeV electrons inside a pulsed-power-driven 13 T magnetic mirror field with a mirror ratio of 2.6. The confinement is diagnosed by measuring the axial and radial losses with magnetic spectrometers. The loss spectra are consistent with $\leq 2.5$ MeV electrons confined in the mirror for $\sim 1$ ns. With a source of $10^{12}$ electron-positron pairs at comparable energies, this magnetic mirror would confine a relativistic pair plasma with Lorentz factor $\gamma \sim 6$ and magnetization $\sigma \sim 40$.

## I. INTRODUCTION

Pair plasma of positrons and electrons behaves drastically different from electron-ion plasmas due to their unity mass ratio.[1] The unity mass ratio alters the waves[2] and unstable modes[3] in a pair plasma. Compared to electron-proton plasma, electron-positron plasma can achieve relativistic and magnetized conditions relevant to energetic astrophysical phenomena with less kinetic energy (1/1000) and weaker magnetic field (1/30).

Relativistic plasmas are found in energetic astrophysical systems abundant in antimatter, such as supernova remnants, gamma-ray bursts, and active galactic nuclei.[4,5] The jets of active galactic nuclei[6] and the magnetospheres of pulsars[7] contain magnetically dominated plasma, where particles may be energized through relativistic magnetic reconnection[8] and kink instabilities.[9] In all these systems, intense non-thermal radiation is thought to result from collisionless shocks formed by magnetic field amplification or generation in Weibel-like instabilities.[10] These Weibel-like instabilities are electromagnetic instabilities in plasmas with momentum anisotropy.[11] Electromagnetic instabilities generally dominate over electrostatic modes (e.g., the two-stream instability) when the plasma has relativistic energies.[12] Which electromagnetic instability dominates the mode spectrum has been the subject of recent theoretical work investigating the effects of ratio of thermal to directed momentum,[13] magnetization,[14] and mass ratio.[12] The unity mass ratio of electron-positron plasma is predicted to lead to $100\times$ faster growth rates than in electron–proton plasma.[12]

Although laboratory electron-positron plasma reduces the particle energy and magnetic field needed to achieve these relativistic and magnetically dominated regimes, the difficulty in producing a large number of positrons and in containing them for long lifetimes has so far precluded their use. In recent decades, much progress has been made in producing reliable positron sources. Positrons with energies of tens to hundreds of eV can be harnessed with continuous fluxes of $10^9$ s$^{-1}$ through the capture of fission neutrons and moderation[15] and in $\mu$s pulses of $1.5 \times 10^5$ positrons with 40 Hz repetition rates through electron-target interactions and moderation at radio frequency accelerator facilities.[16] Short pulse ($\sim$10 ps) laser-target interactions generate large numbers of multi-MeV positrons,[17] up to a $10^{12}$. This favors laser-matter interactions as source for relativistic pair plasma experiments.[5]

A plasma is relativistic when the Lorentz factor is $\gamma > 2$, i.e., the kinetic energy of particles exceeds their rest mass, 511 keV, for electrons and positrons. To study electromagnetic instabilities with these relativistic pairs, it is necessary for the plasma to be dense enough to attenuate electromagnetic waves. This means that the scale length of the plasma must be larger than the skin depth,

$$\lambda_s = c/\sqrt{ne^2/(\varepsilon_0 \gamma m_e)}, \qquad (1)$$

where $c$ is the speed of light, $n$ is the number density of electrons *and* positrons as both respond quickly to perturbations,[2] $e$ is the elementary charge, $m_e$ is the electron (and positron) mass, and $\epsilon_0$ is the permittivity of free space. In addition, the lifetime of the plasma needs to be greater than the instability growth rate $\tau$, which is related to the inverse of the plasma frequency ($\tau \propto 1/\omega_p$).[17] This can be achieved in either beams of pairs dense enough to result in instability growth during the time of flight[18,19] or in less dense magnetically confined pairs.

Here, we focus on developing magnetic traps in order to study magnetized and relativistic pair plasma.[5] A (relativistic) pair plasma is considered magnetically dominated when the magnetization factor—the ratio of magnetic energy density over particle energy density—exceeds unity,

$$\sigma = (B^2/\mu_0)/(n\gamma m_e c^2), \qquad (2)$$

where $B$ is the magnetic field and $\mu_0$ is the permeability of free space. Magnetic configurations can trap pairs with magnetic mirroring effects. This has been successfully demonstrated with lower-energy positrons: a magnetic mirror[20] trapped eV positrons for $> 70$ ms, the magnetic dipole field of a permanent magnet[21] has trapped eV positrons for $> 1$ s, and the dipole field of a levitating superconductor[22] has trapped 100 keV positrons for 100 $\mu$s.

Here, we report on the design of a cm-sized magnetic mirror and on experiments showing the confinement of laser-generated $\leq 2.5$ MeV electrons for a nanosecond in the 13 T mirror field.

## II. MIRROR DESIGN

Inhomogenous magnetic fields exert a force on pairs opposite to the field gradient,[23] $F = -\mu \nabla B$, where $\mu = \gamma m_e u_\perp^2/(2B)$ is the magnetic moment of the particle with perpendicular velocity $u_\perp$. A magnetic mirror traps charged particles between two maxima of the magnetic field. The pitch angle of the particle determines if it is trapped or escapes through the loss cone. Trapped particles complete three motions in a magnetic mirror: they gyrate around the magnetic field lines, they bounce back and forth between the maxima, and they drift azimuthally around the mirror axis.[5] For a simple magnetic mirror generated by two solenoids, the scale length can be approximated as the solenoid radius. When the magnetic field changes slowly compared to the timescale of these motions, i.e., the Larmor radius $r_L = \gamma m u_\perp/(qB)$ of particles is much smaller than the scale length of the magnetic field, trapped particles follow well-constrained paths. They conserve adiabatic invariants and remain outside the loss cone. Yet, conservation of all adiabatic invariants is not a necessary condition to trap particles for several bounce periods.[22] While initially trapped particles exhibiting non-adiabaticity can move into the loss cone and escape, they can remain trapped for several bounce periods. The magneto-inertial fusion discharge system (MIFEDS)[31–34] at the Omega EP short pulse laser facility can store $\sim 500$ J. This is enough to produce a 13 T magnetic field at the center of two 10 mm coils spaced 14 mm apart with a 28 kA current pulse lasting $\mu$s. This geometry results in a mirror ratio of magnetic field at the coil centers to the mirror center of 2.6. These pulsed power coils do not yet achieve the magnetic field strengths to adiabatically trap relativistic electrons and positrons. For example, keeping the Larmor radius of 2.5 MeV electrons and positrons a factor of 10 shorter than a coil radius of 0.5 cm would require a 20 T field in the center of the mirror. In this section, we review the invariants of trapped particle motion in a magnetic mirror, so that we can analyze the motion and trapping of a single simulated relativistic electron in a mirror created by the MIFEDS mirror. Then, we evaluate the confinement properties of the mirror with 10 000 simulated electrons.

In an axisymmetric system, the energy $H$ and the canonical momentum $P$ in the symmetry direction (here, the azimuthal direction $\phi$) are conserved. In addition to these invariants, each of the three motions in the mirror is associated with an adiabatic invariant, which is conserved if the magnetic field varies slowly spatially and temporally over a cycle of the motion. The invariants can be derived from action integrals of the canonical momentum associated with the three motions: gyration perpendicular to the field $P_\perp$, bounce parallel to the field $P_\parallel$, and drift in the azimuthal direction $P_\phi$, along the respective path $dl$.[24] The relativistic adiabatic invariant of the gyromotion $M_r$ differs from the magnetic moment used to calculate the mirror force by an additional factor of gamma, $M_r = \int P_\perp dl = \gamma \mu$. The longitudinal invariant $J$ constrains the bounce motion $u_\parallel$, $J = \int P_\parallel dl = \gamma u_\parallel dl$. In this study, the third invariant $\psi$ can be expressed as the canonical vorticity flux,[25–27] the sum of magnetic flux enclosed by the drift motion and angular momentum $\psi = \int P_\phi dl = \int q_e A_\phi dl + \int \gamma m_e u_\phi dl \sim 2\pi r q_e A_\phi + 2\pi r \gamma m_e u_\phi$, where $r$ is the radial distance of the particle from the mirror axis, $A_\phi$ is the azimuthal magnetic vector potential, and $u_\phi$ is the azimuthal particle velocity. The third invariant does not reduce to solely the magnetic flux because the magnitude of the angular momentum is of the same order. In an axisymmetric system, the third invariant is perfectly conserved even if the electron drifts over strongly varying fields, as the invariant depends directly on the azimuthal canonical momentum.[28]

Integrating the relativistic equation of motion, $d/dt(\gamma m_e \vec{u}) = q\vec{u} \times \vec{B}$ with the Higuera-Cary leapfrog scheme[29] implemented in the AlGeoJ code,[30] reveals that a 2.5 MeV electron is trapped in the MIFEDS mirror field with 5 T in the mirror center and 13 T in the coil centers (mirror ratio 2.6) for several bounce periods despite a path unconstrained by the magnetic moment and the bounce invariant [Fig. 1(a)]. Saitoh et al.[22] showed that such trapped particles with non-conserved first and second adiabatic invariants follow chaotic paths. Strong sensitivity of the chaotic system to initial conditions makes the electron path dependent on the time step. The numerical calculations of the invariants of motion, energy, and canonical momentum exhibit sufficient conservation when the number of time steps exceeds $10^4$ per mean gyration period. Despite the chaotic path, the drift motion is axis encircling [Fig. 1(b)] due to the conservation of the third invariant. As the magnetic flux is not conserved, the electron drift does not lie on a cylindrical surface but makes small displacements off the surface to compensate for the varying angular momentum, enclosing a surface of constant canonical vorticity [Fig. 1(c)].

The MIFEDS coils can be arranged, so that the mirror is centered at target chamber center (TCC) where a target is placed to generate pairs through laser-matter interaction. The trapping of electrons and positrons in this mirror will depend on the pitch angles of perpendicular velocity to parallel velocity with which the particles are injected into the mirror. A previous characterization[35] of electrons and positrons leaving a 1-mm Au target measured an angular distribution with full widths at half maximum (FWHM) of 50°±10° with the beam centered on the rear target normal direction. The divergence can be characterized in terms of the polar angle $\theta$ with respect to the mirror axis (Fig. 2). For a target placed in the center of the mirror with the target normal perpendicular to the mirror axis, the angular distribution of particles would be centered along the target normal, at 90° to the mirror axis.

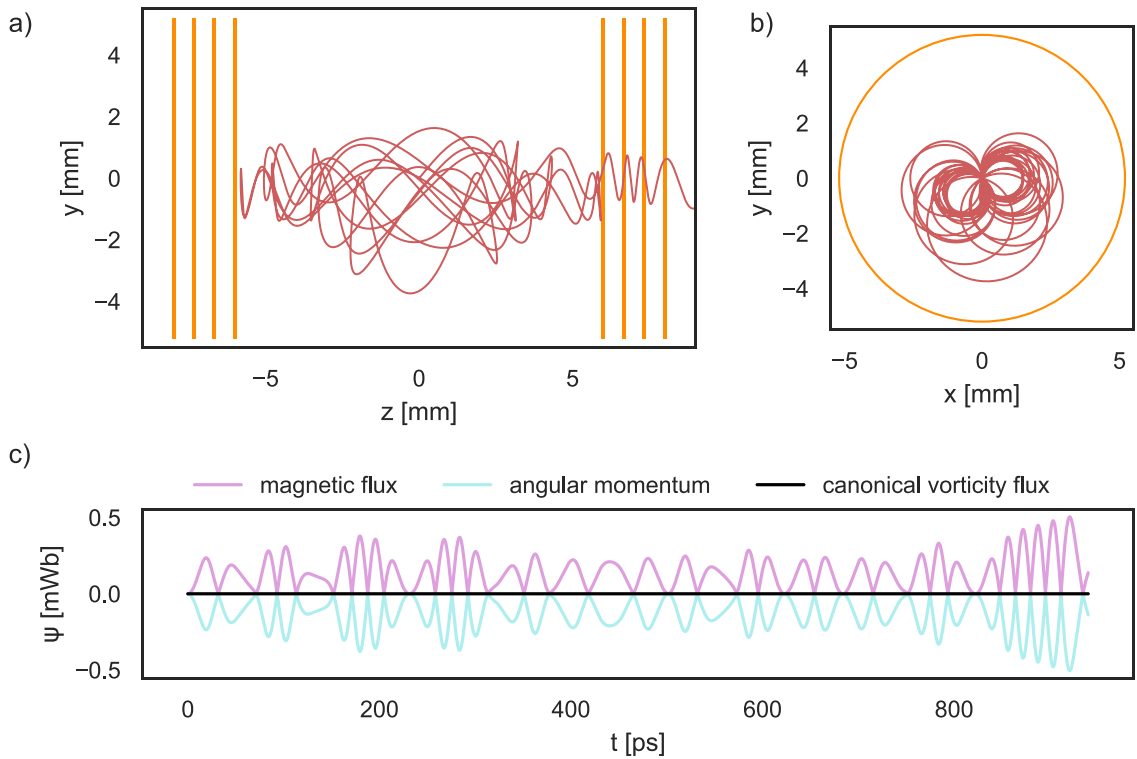

**FIG. 1.** Numerically integrated path of a 2.5 MeV electron launched with pitch angle 51° from the mirror center. The electron is lost after ∼1 ns. (a) View perpendicular to mirror axis shows gyro and bounce motion. (b) View along mirror axis shows drift. (c) Time evolution of the magnetic flux $2\pi r_L A_\phi$, angular momentum $2\pi r_L (\gamma m_e/q_e) u_\phi$, and their sum, the canonical vorticity flux.

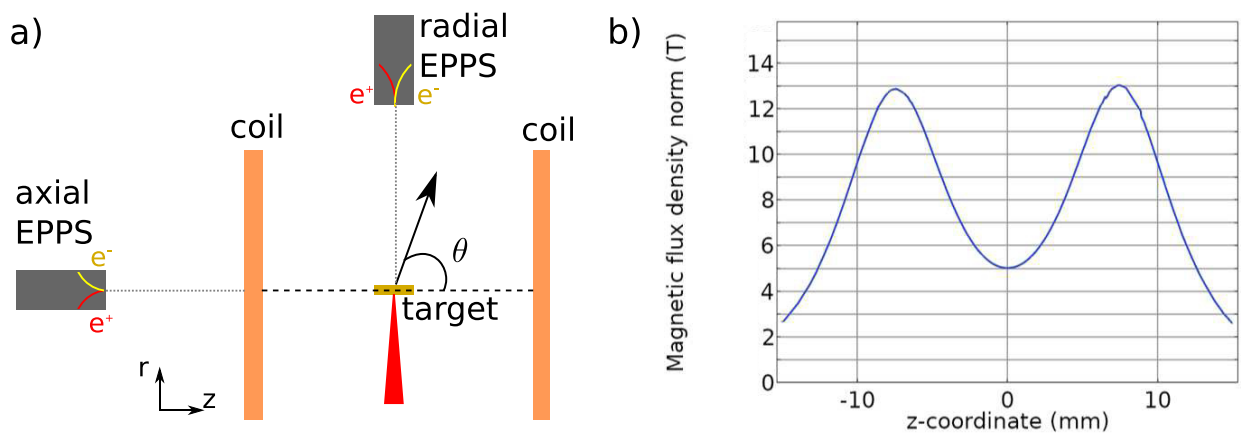

**FIG. 2.** (a) Laser-target interaction generates electrons in mirror field generated by two MIFEDS coils (orange). Target (gold) is centered between coils. Two EPPS magnetic spectrometers measure axial and radial losses (dotted lines) separating electrons (yellow) and positrons (red). The electron divergence is characterized by the polar angle $\theta$ with respect to the mirror axis (dashed line) (not to scale). (b) Magnetic field along mirror axis. $z = 0$ is the mirror center.

To evaluate the confinement properties of the MIFEDS mirror, we numerically integrate the paths of 10 000 electrons with a 2.5 MeV kinetic energy and launched with this angular divergence from the center of the mirror with COMSOL Multiphysics®[36] using the fifth order Dormand-Prince method.[37] Figure 3(a) shows the time evolution of the number of 2.5 MeV electrons in the 13 T mirror (blue). The e-folding time of the number of electrons is ∼2 ns. The losses are due to axial outflows of electrons, including from initially trapped electrons that move non-adiabatically into the loss cone. As the mirror is a closed system, the phase space of electrons can be described with Liouville's theorem, and all trapped electron paths, even chaotic ones, will eventually cross their origin, the target. Collisions with the target scatter the electrons, perhaps shifting their pitch angle into the loss cone. Even with the pessimistic assumption that all target collisions are losses, electrons produced with a 20 μm thick and 0.5 mm diameter target are still trapped for several bounces with e-folding time of 1 ns [Fig. 3(b)].

We can diagnose losses from the mirror with magnetic spectrometers placed outside the mirror. Figure 4 shows the trajectories of 10 000 electrons with energies 2, 3, 5, and 13 MeV in the mirror, without any collective effects. 2 MeV electrons are only lost axially and only for shallow launching angles with respect to the mirror axis, while at higher energies, electrons are also lost radially. At higher energies, an increasing portion of electrons is not trapped at all. Instead, they exit the mirror in less than one gyro-orbit. The trajectories of 3, 5, and 13 MeV electrons exhibit non-axisymmetry, as the electrons sweep a radial loss angle correspondingly to the portion of the last gyro-orbit they complete before leaving the mirror field. There is also a gap in losses at polar angles intersecting the coils. That is, an electron cannot escape by penetrating the coil due to a strong field near the wire surface. Two orthogonal magnetic electron-positron-proton spectrometers (EPPS)[38] can validate these calculated loss spectra. Ratios of the flux F of 10 000 simulated electrons launching from the target with no magnetic field applied and the magnetic field applied $F(B = 13\text{ T})/F(B = 0\text{ T})$ and $F(B = 9\text{ T})/F(B = 0\text{ T})$ quantify the effect of the magnetic mirror without requiring knowledge of the absolute electron numbers (Fig. 5). The magnetic field increases the number of ≤ 2.5 MeV electrons exiting the mirror radially (ratio < 1) and correspondingly increases the number exiting axially (ratio > 1). At higher energies, the effect of the magnetic field becomes negligible (ratio ∼1); however, the ratios do not vary monotonically with energy.

## III. EXPERIMENT SETUP

We conduct experiments at Omega EP to demonstrate the confinement properties of the high field mirror for relativistic electrons. This is because (1) the mirror trapping mechanism does not depend on the sign of charge, and (2) the energy of the positrons from laser-target interactions[35,39] is 5–30 MeV, too high for currently achievable magnetic fields without additional methods to reduce the target-normal sheath acceleration (TNSA)[40] of positrons. The target dimensions, 20 μm thick and 0.5 mm diameter, are optimized for electron production. The short 20 μm thickness of the target limits the production of positrons through the Bethe-Heitler process[41] to levels just above the noise floor. The relativistic electrons are generated by the interaction of an Omega EP wavelength $\lambda = 1054$ nm, 10 ps FWHM pulse[42] with the target. A background shot is taken with no applied magnetic field. The magnetic mirror field is applied for five shots: three with 13 T and two with 9 T at the coil center. The laser energy is $900 \pm 20$ J for the shot without magnetic field and the first shot with 13 T. However, after this first magnetized shot, MIFEDS debris and copper disposition on the laser optics limit the laser energy to $770 \pm 40$ J for all remaining shots. With 80% of the laser energy contained within a $16 \pm 2$ μm radius, the laser energies correspond to intensities of $I = 9 \pm 1 \times 10^{18}$ and $I = 7 \pm 1 \times 10^{18}$ Wcm$^{-2}$, with a laser power contrast of about $10^8$–$10^9$. At such laser conditions, the radially escaping spectrum (Fig. 6) fits a Maxwellian-like exponential with temperature $T = 5.8 \pm 1$ MeV, close to the expected value according to Pukhov scaling,[43] 4.7 MeV, and within previous measured range[39] for the unmagnetized shot laser intensity $9 \times 10^{18}$ Wcm$^{-2}$.

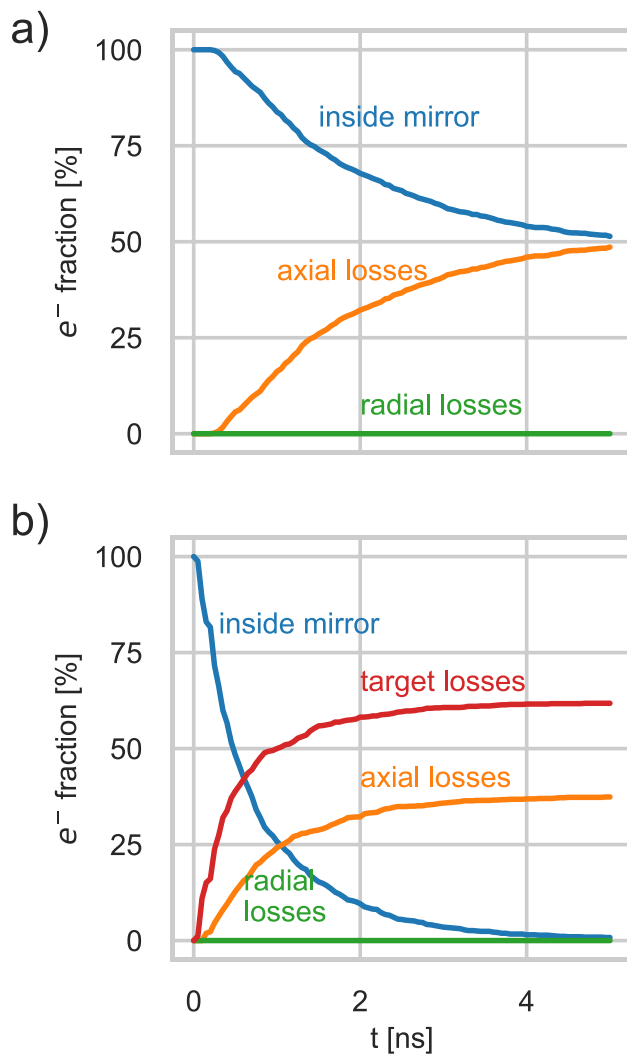

**FIG. 3.** Trapped 2.5 MeV electrons (blue) in the 13 T magnetic mirror. (a) If ignoring the target, the trapping time is limited by axial outflows (red). (b) Under the pessimistic assumption that all collisions with target (green) scatter into the loss cone they further limit the trapping time.

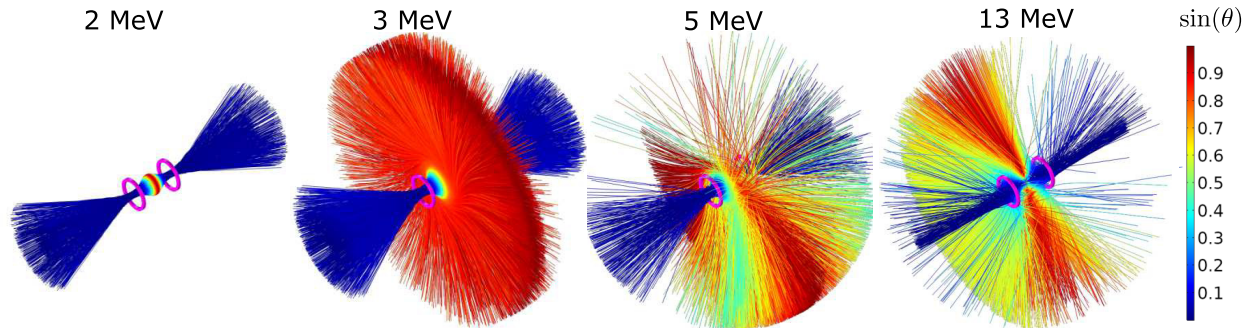

**FIG. 4.** Trajectories of 2, 3, 5, and 13 MeV electrons exiting magnetic mirror formed by two 13 T coils (coils shown in pink). 2 MeV electrons are only lost axially and only for shallow launching angles with respect to the mirror axis, while at higher energies, electrons are also lost radially. 3, 5, and 13 MeV radial electron losses are non-axisymmetric, and the electrons sweep a radial loss angle with increasing energy. The color of the trajectories corresponds to $\sin(\theta)$, where $\theta$ is the polar angle with respect to the mirror axis at which the electron is launched.

## IV. RESULTS AND ANALYSIS

Two electron-positron-proton spectrometers (EPPS) are positioned to sample the electron losses parallel to the MIFEDS magnetic mirror axis and normal to the mirror axis, radial with respect to the mirror geometry (Fig. 2). Each EPPS accepts particles through a $0.95 \times 0.96$ mm$^2$ slit. The radial EPPS is 48 cm, and the axial EPPS is 57 cm from the center of the mirror, which is also the target chamber center. After each shot, the image plates from the two EPPS are scanned after 25 minutes and adjusted for fading with a factor of 0.67 as derived from Tanaka et al.[44] Two super-Gaussians are fitted to the background and signal on each image plate in order to remove the background and determine the projected slit width.[45] Calibrations of the absolute dose[46] and dispersion with measurements in the expected trapped particle energies,[47] 3–15 MeV, provide absolute electron energy spectra. Without the magnetic field, the electrons exit mostly radially, normal to the target surface. The electron energy distribution is determined by the laser-plasma acceleration processes in the preformed plasma on the surface of the target.[48,49] Refluxing of electrons in the target may broaden the energy distribution. The broad angular divergence of electrons leaving the target results in axial losses measured by the EPPS even when no magnetic field is applied. In shots with applied magnetic field, the radial electron loss spectra have three characteristic differences to the baseline no-magnetic-field shot (Fig. 6): (1) the radial EPPS measures a significantly lower number of electrons, (2) the energy spectrum has a steep drop at 3.7 MeV, and (3) followed by a steady increase until 14 MeV (18 MeV) for the 9 T (13 T) field. The energy spectrum of axial losses is shifted upward in shots with a magnetic field, and there are several narrow spikes with energies below 5 MeV followed by a broad spike at 10 MeV (13 MeV) for 9 T (13 T). The spikes in the axial electron spectra can be explained by magnetic lensing effects.[50] Each solenoid acts as a lens for charged particles with a focal length $f$ that can be approximated by $f \sim 3.4 \rho_{ce}^2/a$, where $\rho_{ce} = \gamma m_e u/(q_e B_0)$ is the electron gyroradius, where $B_0$ is the magnetic field at the coil center, $q_e$ is the charge, and $a$ is the coil diameter. The energy of the broad peaks in the axial spectrum corresponds to the electron energy, for which the focal length is the coil-to-target distance 7.5 mm, i.e., the energy of electrons that should be collimated.[51] The narrower spikes in the measured axial loss spectra below 5 MeV have been noted in magnetic focusing experiments[41] and correspond to complicated trajectories of specific energy electrons, which focus and subsequently re-collimate upon exiting the coil. There is significant variation in the spectra between shots; this may be related to variation in the laser parameters, tilt off the mirror axis (3°–5°) in the construction of the coils,[51] and shot-to-shot variation in the magnitude of the magnetic field. The radial spectra have a strong dependence on the magnetic field magnitude because of the large azimuthal rotation introduced by the field. Despite this variation, the characteristics described here are recognizable across all shots. To relate the measured losses to the calculated particle trajectories, we compare the numerical electron flux $F(B = 13\ T)/F(B = 0\ T)$ from the calculated trajectories to the ratios of the measured spectra (Fig. 5). The simulated flux ratios are consistent with the characteristics observed in the measured fluxes: (1) the radial flux is significantly reduced and the axial flux is correspondingly enhanced, (2) the radial flux sharply drops off for energies above 3.7 MeV, and (3) followed by a steady increase until 18 MeV for the 13 T. The reduction in the radially lost electrons and corresponding increase in axially lost electrons agrees with the path integrations of trapped electrons lost along the axial loss cones.

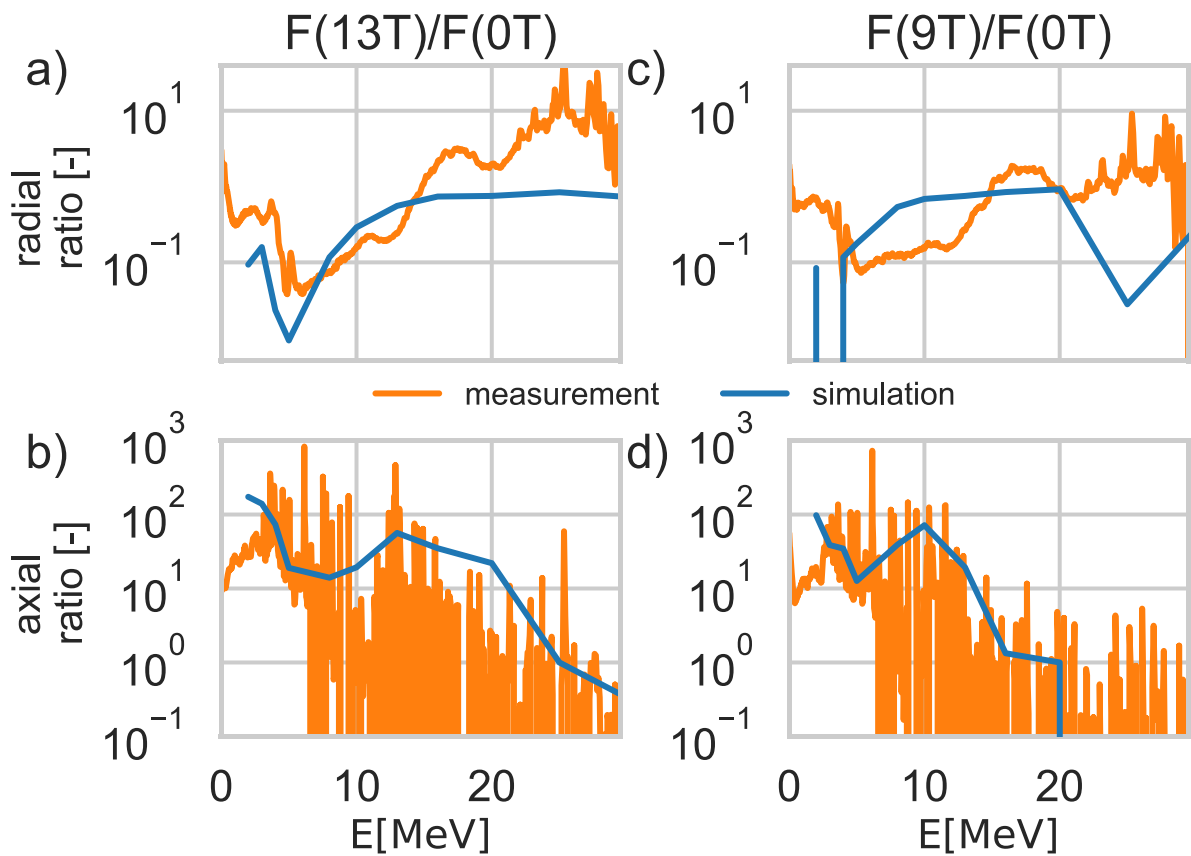

**FIG. 5.** Ratio of $B = 13$ T to $B = 0$ T and $B = 9$ T to $B = 0$ T simulated radial (a and c) and axial (b and d) electron loss fluxes (blue) plotted together with respective ratio of measured loss spectra (orange).

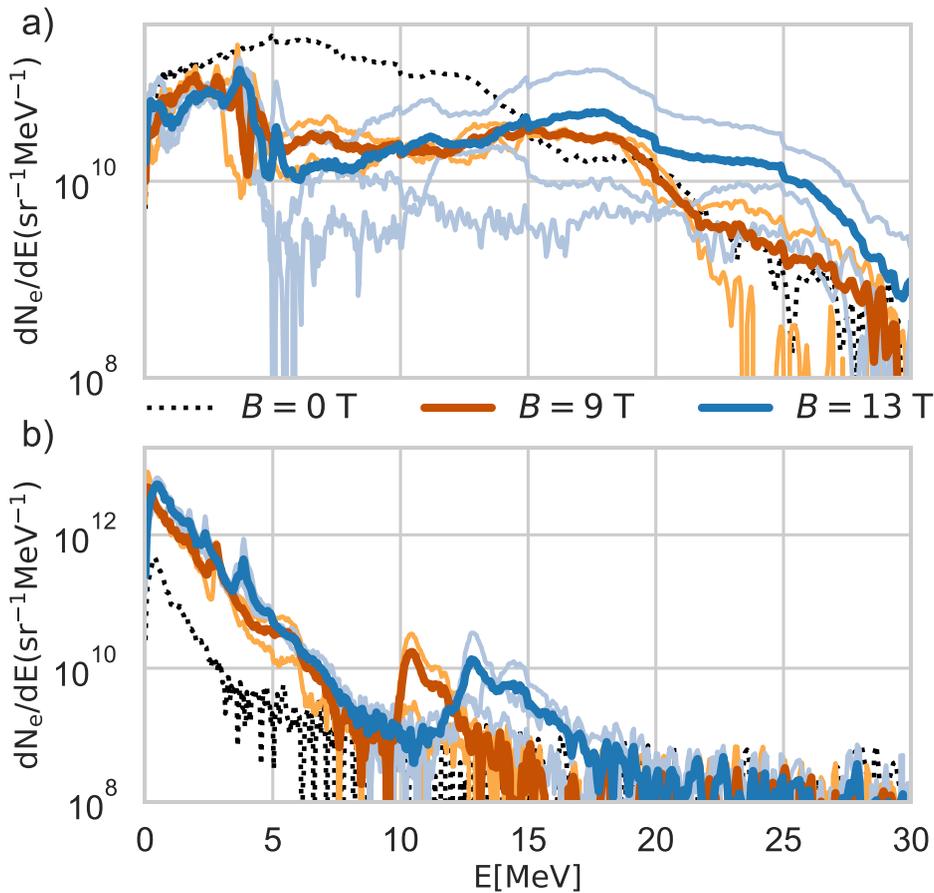

**FIG. 6.** EPPS measured spectra of escaping electrons. Spectra of (a) radially and (b) axially escaping electrons with no magnetic field (dashed back), three shots with 13 T magnetic field at the coil centers (light blue; average of shots thick blue), and two shots with 9 T magnetic field at coil centers (light orange; average of shots thick dark orange).

As the mirror force is independent of the sign of the charge, this mirror should be able to confine positrons and electrons simultaneously. Positrons and electrons could be generated outside of the mirror with thicker targets optimized for positron production. The pairs could then be magnetically transported to the mirror with a solenoid acting as collimating lens. The lens would select particle energy, and if tilted also select charge,[51] so that a unity charge ratio pair beam could be injected into the mirror. A thin foil placed inside the mirror could scatter the injected pairs out of the loss cone.[5] The coils could also be arranged in a cusp configuration with a magnetic null at the center; this may simplify injection from an outside target as the magnetic null would allow pairs to transition non-adiabatically to trapped particles.[52]

With 1 mm thick targets, up to $10^{12}$ positrons have been produced.[17] In the mirror described here, this would result in a density of $n \sim 10^{12}$ cm$^{-3}$. The magnetization would be $\sigma \sim 40$, $B = 5$ T at the mirror center and confined energies of 2.5 MeV ($\gamma \sim 6$) in Eq. (2). This density $n = 10^{12}$ cm$^{-3}$ and relativistic Lorentz factor $\gamma = 6$ would correspond to skin depths of 0.9 cm [Eq. (1)], being on the same order as the coil diameter and separation, characteristics of the mirror geometry. As the Debye length approaches the skin depth in relativistic regimes, a pair plasma of these densities would approach the conditions for supporting electrostatic and electromagnetic instabilities.[5]

## V. CONCLUSION

We have trapped $\leq 2.5$ MeV electrons for a nanosecond in a 13 T magnetic mirror field and, thereby, demonstrated that magnetic mirror coils driven by the existing MIFEDS power supplies can trap relativistic electrons and positrons. This can be achieved without the conservation of the magnetic moment and bounce invariant, resulting in chaotic but nonetheless trapped particle paths. Measurements of the axial and radial losses from the magnetic mirror are consistent with numerically calculated electron losses and confinement of $\leq 2.5$ MeV electrons for several bounce periods. Future experiments could use a high yield positron source; the largest positron yield achieved with intense laser-matter interaction[17] is $10^{12}$. Trapping this yield in the mirror discussed here would produce a pair plasma approaching the plasma conditions for supporting electrostatic and electromagnetic instabilities. The pairs would be relativistic with $\gamma \sim 6$ and magnetized with $\sigma \sim 40$. Increases in the magnetic field could increase the energy of pairs that can be trapped. For example, the upgrade of the pulsed power supply at OMEGA from MIFEDS-2 to MIFEDS-3 will increase the stored electric energy from 200 J to 2 kJ, enabling higher currents and corresponding magnetic fields to be driven in the coils.[33]

## ACKNOWLEDGMENTS

We gratefully acknowledge the support of the Omega EP facility during these Laboratory Basic Science experiments and helpful discussions about adiabatic invariants for relativistic trapped particles with Eve V. Stenson, Stefan Nissl, Matthew R. Stoneking, and Thomas Sunn Pedersen. This work was performed under the auspices of the U.S. Department of Energy by Lawrence Livermore National Laboratory under Contract DE-AC52-07NA27344 and was supported by the LLNL-LDRD Program under Project No. 20-LW-021. L.W. is supported by the National Science Foundation under Award Number 1751462. J.v.d.L. is supported by the Alexander von Humboldt Foundation, LLNL-JRNL-822631.

This document was prepared as an account of work sponsored by an agency of the United States government. Neither the United States government nor Lawrence Livermore National Security, LLC, nor any of their employees makes any warranty, expressed or implied, or assumes any legal liability or responsibility for the accuracy, completeness, or usefulness of any information, apparatus, product, or process disclosed, or represents that its use would not infringe privately owned rights. Reference herein to any specific commercial product, process, or service by trade name, trademark, manufacturer, or otherwise does not necessarily constitute or imply its endorsement, recommendation, or favoring by the United States government or Lawrence Livermore National Security, LLC. The views and opinions of authors expressed herein do not necessarily state or reflect those of the United States government or Lawrence Livermore National Security, LLC, and shall not be used for advertising or product endorsement purposes.

The authors have no conflicts to disclose.

## DATA AVAILABILITY

The data that support the findings of this study are available from the corresponding author upon reasonable request.

Physics of Plasmas　　　ARTICLE　　　scitation.org/journal/php
[12] A. Bret, A. Stockem, F. Fiuza, C. Ruyer, L. Gremillet, R. Narayan, and L. O. Silva, "Collisionless shock formation, spontaneous electromagnetic fluctuations, and streaming instabilities," Phys. Plasmas 20, 042102 (2013).

[13] A. Stockem, F. Fiuza, A. Bret, R. A. Fonseca, and L. O. Silva, "Exploring the nature of collisionless shocks under laboratory conditions," Sci. Rep. 4, 3934 (2014).

[14] A. Bret, "Hierarchy of instabilities for two counter-streaming magnetized pair beams," Phys. Plasmas 23, 062122 (2016).

[15] C. Hugenschmidt, C. Piochacz, M. Reiner, and K. Schreckenbach, "The NEPOMUC upgrade and advanced positron beam experiments," New J. Phys. 14, 055027 (2012).

[16] H. Higaki, K. Michishio, K. Hashidate, A. Ishida, and N. Oshima, "Accumulation of LINAC based low energy positrons in a buffer gas trap," Appl. Phys. Express 13, 066003 (2020).

[17] H. Chen, F. Fiuza, A. Link, A. Hazi, M. Hill, D. Hoarty, S. James, S. Kerr, D. Meyerhofer, J. Myatt, J. Park, Y. Sentoku, and G. Williams, "Scaling the yield of laser-driven electron-positron jets to laboratory astrophysical applications," Phys. Rev. Lett. 114, 215001 (2015).

[18] E. Liang, T. Clarke, A. Henderson, W. Fu, W. Lo, D. Taylor, P. Chaguine, S. Zhou, Y. Hua, X. Cen, X. Wang, J. Kao, H. Hasson, G. Dyer, K. Serratto, N. Riley, M. Donovan, and T. Ditmire, "High $e^+/e^-$ ratio dense pair creation with $10^{21}$ w.cm$^{-2}$ laser irradiating solid targets," Sci. Rep. 5, 13968 (2015).

[19] G. Sarri, K. Poder, J. M. Cole, W. Schumaker, A. D. Piazza, B. Reville, T. Dzelzainis, D. Doria, L. A. Gizzi, G. Grittani, S. Kar, C. H. Keitel, K. Krushelnick, S. Kuschel, S. P. D. Mangles, Z. Najmudin, N. Shukla, L. O. Silva, D. Symes, A. G. R. Thomas, M. Vargas, J. Vieira, and M. Zepf, "Generation of neutral and high-density electron–positron pair plasmas in the laboratory," Nat. Commun. 6, 6747 (2015).

[20] H. Higaki, C. Kaga, K. Fukushima, H. Okamoto, Y. Nagata, Y. Kanai, and Y. Yamazaki, "Simultaneous confinement of low-energy electrons and positrons in a compact magnetic mirror trap," New J. Phys. 19, 023016 (2017).

[21] J. Horn-Stanja, S. Nißl, U. Hergenhahn, T. S. Pedersen, H. Saitoh, E. Stenson, M. Dickmann, C. Hugenschmidt, M. Singer, M. Stoneking, and J. Danielson, "Confinement of positrons exceeding 1 s in a supported magnetic dipole trap," Phys. Rev. Lett. 121, 235003 (2018).

[22] H. Saitoh, Z. Yoshida, Y. Yano, M. Nishiura, Y. Kawazura, J. Horn-Stanja, and T. S. Pedersen, "Chaos of energetic positron orbits in a dipole magnetic field and its potential application to a new injection scheme," Phys. Rev. E 94, 043203 (2016).

[23] D. R. Nicholson, Introduction to Plasma Theory (Wiley & Sons, New York, 1983).

[24] T. G. Northrop, "Adiabatic charged-particle motion," Rev. Geophys. 1, 283, https://doi.org/10.1029/RG001i003p00283 (1963).

[25] S. You, "A field theory approach to the evolution of canonical helicity and energy," Phys. Plasmas 23, 072108 (2016).

[26] E. S. Lavine and S. You, "The topology of canonical flux tubes in flared jet geometry," Astrophys. J. 835, 89 (2017).

[27] J. von der Linden, J. Sears, T. Intrator, and S. You, "Measurements of the canonical helicity of a gyrating kink," Phys. Rev. Lett. 121, 035001 (2018).

[28] B. P. M, Fundamentals of Plasma Physics (Cambridge University Press, 2004).

[29] A. V. Higuera and J. R. Cary, "Structure-preserving second-order integration of relativistic charged particle trajectories in electromagnetic fields," Phys. Plasmas 24, 052104 (2017).

[30] S. Nißl, "Numerical investigation into injection and confinement of single particles in a magnetic dipole trap," Master's thesis (Technical University of Munich, 2018).

[31] G. Fiksel, R. Backhus, D. H. Barnak, P.-Y. Chang, J. R. Davies, D. Jacobs-Perkins, P. McNally, R. B. Spielman, E. Viges, and R. Betti, "Inductively coupled 30 T magnetic field platform for magnetized high-energy-density plasma studies," Rev. Sci. Instrum. 89, 084703 (2018).

[32] D. H. Barnak, J. R. Davies, G. Fiksel, P.-Y. Chang, E. Zabir, and R. Betti, "Increasing the magnetic-field capability of the magneto-inertial fusion electrical discharge system using an inductively coupled coil," Rev. Sci. Instrum. 89, 033501 (2018).

[33] R. Shapovalov, G. Brent, R. Moshier, M. Shoup, R. Spielman, and P.-A. Gourdain, "Design of 30-T pulsed magnetic field generator for magnetized high-energy-density plasma experiments," Phys. Rev. Accel. Beams 22, 080401 (2019).

[34] J. Peebles, J. R. Davies, G. Brent, D. Mastrosimone, D. W. Jacobs-Perkins, G. Fiksel, M. J. Shoup III, T. Lewis, G. Gates, P. A. Gourdain, R. Shapovalov, R. Moshier, T. Burgett, and R. Betti, "Current capabilities of the MIFEDS system," paper presented at the Meeting on Magnetic Fields in Laser Plasmas, Rochester, NY, 23–24 April 2018.

[35] H. Chen, S. C. Wilks, D. D. Meyerhofer, J. Bonlie, C. D. Chen, S. N. Chen, C. Courtois, L. Elberson, G. Gregori, W. Kruer, O. Landoas, J. Mithen, J. Myatt, C. D. Murphy, P. Nilson, D. Price, M. Schneider, R. Shepherd, C. Stoeckl, M. Tabak, R. Tommasini, and P. Beiersdorfer, "Relativistic quasimonoenergetic positron jets from intense laser-solid interactions," Phys. Rev. Lett. 105, 015003 (2010).

[36] COMSOL Multiphysics® v. 5.5, COMSOL AB, Stockholm, Sweden.

[37] J. Dormand and P. Prince, "A family of embedded runge-kutta formulae," J. Comput. Appl. Math. 6, 19–26 (1980).

[38] H. Chen, A. J. Link, R. van Maren, P. K. Patel, R. Shepherd, S. C. Wilks, and P. Beiersdorfer, "High performance compact magnetic spectrometers for energetic ion and electron measurement in ultraintense short pulse laser solid interactions," Rev. Sci. Instrum. 79, 10E533 (2008).

[39] H. Chen, A. Link, Y. Sentoku, P. Audebert, F. Fiuza, A. Hazi, R. F. Heeter, M. Hill, L. Hobbs, A. J. Kemp, G. E. Kemp, S. Kerr, D. D. Meyerhofer, J. Myatt, S. R. Nagel, J. Park, R. Tommasini, and G. J. Williams, "The scaling of electron and positron generation in intense laser-solid interactions," Phys. Plasmas 22, 056705 (2015).

[40] S. C. Wilks, A. B. Langdon, T. E. Cowan, M. Roth, M. Singh, S. Hatchett, M. H. Key, D. Pennington, A. MacKinnon, and R. A. Snavely, "Energetic proton generation in ultra-intense laser–solid interactions," Phys. Plasmas 8, 542–549 (2001).

[41] G. J. Williams, D. Barnak, G. Fiksel, A. Hazi, S. Kerr, C. Krauland, A. Link, M. J.-E. Manuel, S. R. Nagel, J. Park, J. Peebles, B. B. Pollock, F. N. Beg, R. Betti, and H. Chen, "Target material dependence of positron generation from high intensity laser-matter interactions," Phys. Plasmas 23, 123109 (2016).

[42] L. Waxer, D. Maywar, J. Kelly, T. Kessler, B. Kruschwitz, S. Loucks, R. McCrory, D. Meyerhofer, S. Morse, C. Stoeckl, and J. Zuegel, "High-energy petawatt capability for the omega laser," Opt. Photonics News 16, 30 (2005).

[43] A. Pukhov, Z.-M. Sheng, and J. Meyer-ter-Vehn, "Particle acceleration in relativistic laser channels," Phys. Plasmas 6, 2847–2854 (1999).

[44] K. A. Tanaka, T. Yabuuchi, T. Sato, R. Kodama, Y. Kitagawa, T. Takahashi, T. Ikeda, Y. Honda, and S. Okuda, "Calibration of imaging plate for high energy electron spectrometer," Rev. Sci. Instrum. 76, 013507 (2005).

[45] J. Park, "The effects of preformed plasmas on the directionality of the relativistic electron beam," Ph.D. thesis (University of California, 2015).

[46] H. Chen, N. L. Back, T. Bartal, F. N. Beg, D. C. Eder, A. J. Link, A. G. MacPhee, Y. Ping, P. M. Song, A. Throop, and L. Van Woerkom, "Absolute calibration of image plates for electrons at energy between 100 KeV and 4 MeV," Rev. Sci. Instrum. 79, 033301 (2008).

[47] J. von der Linden, J. Ramos-Méndez, B. Faddegon, D. Massin, G. Fiksel, J. P. Holder, L. Willingale, J. Peebles, M. R. Edwards, and H. Chen, "Dispersion calibration for the National Ignition Facility electron–positron–proton spectrometers for intense laser matter interactions," Rev. Sci. Instrum. 92, 033516 (2021).

[48] S. C. Wilks, W. L. Kruer, M. Tabak, and A. B. Langdon, "Absorption of ultra-intense laser pulses," Phys. Rev. Lett. 69, 1383–1386 (1992).

[49] A. Kemp, F. Fiuza, A. Debayle, T. Johzaki, W. Mori, P. Patel, Y. Sentoku, and L. Silva, "Laser–plasma interactions for fast ignition," Nucl. Fusion 54, 054002 (2014).

[50] H. Chen, G. Fiksel, D. Barnak, P.-Y. Chang, R. F. Heeter, A. Link, and D. D. Meyerhofer, "Magnetic collimation of relativistic positrons and electrons from high intensity laser–matter interactions," Phys. Plasmas 21, 040703 (2014).

[51] J. L. Peebles, G. Fiksel, M. R. Edwards, J. von der Linden, L. Willingale, D. Mastrosimone, and H. Chen, "Magnetically collimated relativistic charge-neutral electron-positron beams from high-power lasers," Phys. Plasmas 28, 074501 (2021).

[52] J. Park, N. A. Krall, P. E. Sieck, D. T. Offermann, M. Skillicorn, A. Sanchez, K. Davis, E. Alderson, and G. Lapenta, "High-energy electron confinement in a magnetic cusp configuration," Phys. Rev. X 5, 021024 (2015).

Phys. Plasmas 28, 092508 (2021); doi: 10.1063/5.0057582　　　28, 092508-8
© Author(s) 2021